\documentclass[%
 reprint,
superscriptaddress,
 amsmath,amssymb,
 aps,
]{revtex4-2}

\usepackage{graphicx}
\usepackage{dcolumn}
\usepackage{bm}
\usepackage{xcolor}
\usepackage{color}
\usepackage[utf8x]{inputenc}
\usepackage[normalem]{ulem}
\usepackage{xurl}
\usepackage[colorlinks = true,
            linkcolor = blue,
            urlcolor  = blue,
            citecolor = blue,
            anchorcolor = blue]{hyperref}

\begin{document}
\title{
A Unsupervised Framework for Identifying Diverse Quantum Phase Transitions Using Classical Shadow Tomography
}

\author{Chi-Ting Ho}
\affiliation{Physics Department, National Tsing Hua University, Hsinchu 30013, Taiwan}
\affiliation{Center for Theory and Computation, National Tsing Hua University, Hsinchu 30013, Taiwan}

\author{Daw-Wei Wang}
\affiliation{Physics Department, National Tsing Hua University, Hsinchu 30013, Taiwan}
\affiliation{Center for Theory and Computation, National Tsing Hua University, Hsinchu 30013, Taiwan}
\affiliation{Center for Quantum Technology, National Tsing Hua University, Hsinchu 30013, Taiwan}

\date{\today} 

\begin{abstract}
{
We provide a general machine learning methodology that integrates classical shadow representations with unsupervised principal component analysis (PCA) to explore various quantum phase transitions. By sampling spin configurations from random Pauli measurements, our approach can effectively analyze hidden statistical patterns in the data, thereby capturing the distinct signatures of quantum criticality through their fluctuations. We benchmark this approach across various spin-1/2 systems, including the 1D XZX cluster-Ising model, the 1D bond-alternating XXZ model, the 2D transverse-field Ising model, and the 2D Kitaev honeycomb model. We show that PCA not only reliably detects and distinguishes both symmetry-breaking and topological transitions, but also enables their qualitative classification based on characteristic fluctuation patterns. Our data-driven approach does not require any knowledge of the Hamiltonian or explicit order parameters, and can therefore be a general and applicable tool for probing new quantum phases.
}
\end{abstract}
\maketitle

\section{Introduction}
\label{sec: introduction}
Exploring quantum phase transitions (QPTs) in many-body systems remains a fundamental challenge in condensed matter physics. Conventional approaches to phase identification typically rely on measuring specific order parameters or evaluating relevant physical quantities from quantum states. However, exotic quantum phases beyond the Landau symmetry-breaking paradigm, such as topologically ordered phases and quantum spin liquids \cite{intro_QSL_1,intro_QSL_2}, are characterized by non-local order parameters and long-range entanglement, necessitating alternative theoretical and experimental frameworks for their accurate description \cite{book_Bernevig}. The experimental detection of these non-local order parameters remains a significant challenge, as conventional measurement techniques often fail to capture the underlying long-range entanglement and topological properties. 

Recent developments in quantum simulations have introduced randomized measurement protocols \cite{intro_randomized-measurement_1, intro_randomized-measurement_2, intro_randomized-measurement_3} and the classical shadow framework \cite{intro_classical-shadow_1, intro_classical-shadow_2}, which enable efficient estimation of physical observables from a limited set of measurement outcomes. By compressing quantum state information into compact classical representations, classical shadows render the estimation of various observables without requiring full quantum state tomography. This makes them particularly advantageous for probing non-local correlations and complex quantum properties.

In parallel, machine learning techniques offer a complementary and potential route for revealing hidden structures and critical phenomena in many-body physics. Although supervised learning approaches \cite{paper_SL_1, paper_SL_2, paper_SL_3, paper_SL_4, paper_SL_5, paper_SL_6, paper_SL_7, paper_SL_8} require labeled data and thus the prior knowledge of these phases, unsupervised \cite{paper_USL_1, paper_USL_2, paper_USL_3, paper_USL_4, paper_USL_5, paper_USL_6, paper_USL_7, paper_USL_8, paper_USL_9, paper_USL_10, paper_USL_11, paper_USL_12, paper_USL_13, paper_USL_14, paper_USL_15, paper_USL_16} and self-supervised learning \cite{paper_Ho_1, paper_Ho_2} techniques have shown great promise in extracting informative features directly from raw data (either from experimental measurements or numerical simulations). These two approaches enable phase identification and boundary detection without the predefined order parameters or model-specific assumptions, making them particularly attractive for studying complex or previously unexplored quantum systems from experimental data. Nevertheless, prior approaches rely on specific measurements and may not be universally applicable to alternative experiments or systems. It remains important to explore the development of a simpler and more generalized methodology for future research advancements.

In this work, we propose an integrated methodology based on classical shadow representations and unsupervised principal component analysis (PCA) \cite{paper_PCA}. Different from the conventional unsupervised clustering methods, which typically aim to classify quantum phases by clustering the similar macroscopic properties of the ground states within the same phases, our PCA-based approach considers a quantum phase transition based on characterizing the signatures of quantum fluctuations near the critical point. We show that this approach could extract dominant features in the data that not only reveal the presence of quantum phase transitions but also differentiate between distinct types of phase transitions without requiring prior knowledge of the underlying Hamiltonian. 

More specifically, we apply our approach to various spin-1/2 models, including the 1D XZX cluster-Ising model \cite{paper_cluster-ising_1, paper_cluster-ising_2}, the 1D bond-alternating XXZ model \cite{paper_alternating-xxz_1}, the 2D transverse-field Ising model on a square lattice, and the 2D Kitaev model on a honeycomb lattice \cite{paper_2d-kitaev}. By systematically analyzing the statistical fluctuations captured from the randomized measurement of \textit{in situ} spin configurations, we find that the pattern of leading principal components, obtained from PCA, can reflect the properties of underlying quantum fluctuation as the system parameter varies. We demonstrate that our approach can identify both symmetry-breaking and topological phase transitions, and even distinguish them easily. This highlights a key advantage of our approach, and can be easily applied to other strongly correlated systems with quantum spin 1/2.

This paper is organized as follows. In Sec.~\ref{sec: method}, we briefly review the classical shadow framework and describe how our PCA-based approach identifies a quantum phase transition, using the 1D transverse-field Ising chain. In Sec.~\ref{sec: identification-topological}, we apply the method to the 1D XZX cluster-Ising chain and the 1D bond-alternating XXZ chain to demonstrate its ability to detect topological phase transitions. In Sec.~\ref{sec: 2d-quantum}, we show the applicability for two-dimensional quantum systems, including the 2D transverse-field Ising model and the 2D Kitaev honeycomb model. In Sec.~\ref{sec: discussion}, we discuss why our approach can differentiate symmetry-breaking and topological phase transitions. Finally, we summarize our conclusions and outlook in Sec.~\ref{sec: summary}.

\section{Method}
\label{sec: method}
\subsection{Classical shadow framework for characterizing quantum systems}
\label{sec: classical-shadow}

We begin by introducing the classical shadow framework, which provides an efficient and scalable method for characterizing quantum states \cite{intro_classical-shadow_1, intro_classical-shadow_2}. Instead of performing full quantum state tomography, which is infeasible for large systems due to the exponential growth of the Hilbert space, the classical shadow protocol allows one to estimate physical observables directly from randomized measurements. In this framework, the quantum state is repeatedly measured in randomly chosen bases, typically sampled from unitary ensembles such as the Clifford or Pauli group. Each measurement outcome yields an in-situ spin configuration that captures partial information about the state. By aggregating a sufficiently large number of such spin configurations, one can accurately reconstruct various physical properties of the underlying quantum state.

For concreteness, consider a quantum state in the spin-1/2 system of size $L$. The measurement is performed on a randomly chosen Pauli basis at each site, which yields a spin configuration $S=\{s_1,\cdots,s_L\}$, where $s_i\in\{\pm X,\pm Y,\pm Z\}$ denotes the measured result at site $i$. Repeating this randomized measurement procedure $N$ times produces a collection of spin configurations, which serves as the raw data for reconstructing an approximation to the density matrix:
\begin{equation}
    \rho \approx \frac{1}{N}
    \sum_{n=1}^{N} 
    \bigotimes_{i=1}^{L}
    (3|s^{(n)}_{i}\rangle \langle s^{(n)}_{i}|-I), 
    \label{eqn: classical-shadow}
\end{equation}
where $s^{(n)}_{i}\in\{\pm X, \pm Y, \pm Z\}$ denotes the outcome at site $i$ in the $n$-th measured outcome, and $I$ is the $2\times 2$ identity matrix. Note that each single-site tensor has unit trace, i.e., $\text{Tr}(3|s^{(n)}_{i}\rangle \langle s^{(n)}_{i}|-I)=1$. This ensures that the global reconstruction preserves normalization and enables the efficient estimation of observables, without requiring an exponentially large number of measurements in system size. 

Importantly, the use of randomized Pauli measurements can capture the global information of the quantum state, making the classical shadow framework particularly advantageous for probing non-local properties and exploring exotic quantum phases. In the following, we will describe how the signatures of phase transition can be extracted from a data-driven analysis perspective.

\subsection{PCA-based approach for exploring quantum phase transition}
\label{sec: machine-learning}

We first clarify the task of exploring quantum phase transitions. Consider a spin-1/2 system with an effective Hamiltonian $H(g) = H_{0} + gH_{1}$, where $H_{0}$ and $H_{1}$ are two non-commuting terms, and $g$ is a control parameter. When $g$ is tuned across the critical point $g_{c}$, the competition between $H_0$ and $H_1$ becomes significant, giving rise to a qualitative change in the ground state and a quantum phase transition with enhanced quantum fluctuation \cite{book_sachdev}. Conventionally, one may use the averaged experimental quantities and employ a supervised or unsupervised learning method to identify the phase transition. The basic idea of these scenarios is to treat the experimental data of different control parameters as different objects and classify or cluster them based on the distinct macroscopic properties of ground states deep within each phase. The phase boundaries are then identified when the machine learning grouping or labeling of objects changes.

However, these strategies rely on the presumption that obtained clusters in feature space are clearly separated. If the underlying phase structure does not appear as distinct clusters but instead as gradual changes, it may be very difficult (or even impossible) to accurately determine phase boundaries. This limitation is crucial, especially in systems with topological transitions, where no local order parameter exists, and hence the subtle signatures of different phases are not easily captured. 

As a result, our primary interest lies in the signatures of quantum fluctuation near phase boundaries. Even without explicit knowledge of $H_0$ and $H_1$ (as is often the case in realistic experimental settings), it is reasonable to expect that such critical behavior would still manifest itself in the statistical patterns of the \textit{in situ} measurement. In particular, enhanced quantum fluctuation near criticality is anticipated to increase the diversity of classical shadow spin configurations, in contrast to more uniform patterns found deep within a given phase. However, it should be noted that statistical variation may also arise from the randomized choice of measured Pauli bases. This inherent randomness contributes additional fluctuations to the \textit{in situ} data, which could mix with true quantum fluctuations and thus complicate their separation.

To systematically capture the underlying quantum fluctuation from the overall statistical patterns of the \textit{in situ} data, we employ principal component analysis (PCA) \cite{paper_PCA}, a widely used unsupervised learning technique for identifying dominant patterns in high-dimensional data. The primary objective of PCA is to seek a set of orthonormal basis vectors, known as principal components, that maximize the variance in the dataset. This is achieved by computing the covariance matrix (denoted by C) of the dataset:
\begin{equation}
    \begin{aligned}
    C 
    &= \frac{1}{N}\sum_{n=1}^{N}
    ({\bf x}^{(n)}-\bar{\bf x})({\bf x}^{(n)}-\bar{\bf x})^{T} \\
    &=\frac{1}{N}\sum_{n=1}^{N}
    {\bf x}^{(n)}({\bf x}^{(n)})^{T} - \bar{\bf x}\bar{\bf x}^{T},
    \label{eqn: covariance}
    \end{aligned}
\end{equation}
where $N$ is the number of \textit{in situ} spin configurations obtained by randomized measurement (we take $N=2000$ throughout in this paper), ${\bf x}^{(n)}$ is the $n$-th spin configuration represented as a high-dimensional vector,  and $\bar{\bf x}$ is the mean vector of ${\bf x}^{(n)}$. The eigenvectors of $C$ define the directions of the principal components, and the corresponding eigenvalues indicate the variance computed by each principal component in the dataset. Here we encode each measured outcome at site $i$ ($s_i\in\{\pm X, \pm Y, \pm Z\}$) as a three-dimensional Euclidean vector in regular form: $\pm X =(\pm 1, 0, 0)^T$, $\pm Y =(0, \pm 1, 0)^T$ and $\pm Z =(0, 0, \pm 1)^T$. As a result, a spin configuration $S$ corresponds to a sequence of such vectors across all sites in the system, resulting in a vector of length $3L$ for a system of size $L$. 

In this representation, the covariance matrix can be viewed as consisting of $L \times L$ blocks, with each block being a $3 \times 3$ matrix. Interestingly, according to the classical shadow formula shown in Eq.~(\ref{eqn: classical-shadow}), the elements of each block can be approximated as follows:
\begin{equation}
    c^{(ij)}_{\alpha\beta} \approx 
    \begin{cases}
        \frac{1}{3} \delta_{\alpha\beta}
        (1-\langle \sigma^{\alpha}_i \rangle^2), 
        & \text{if $i=j$}\\
        \frac{1}{9} 
        (\langle \sigma^{\alpha}_i\sigma^{\beta}_j \rangle 
        - \langle \sigma^{\alpha}_i \rangle \langle \sigma^{\beta}_j \rangle),
        & \text{otherwise}\\
    \end{cases}
    \label{eqn: block}
\end{equation}
Here, $c^{(ij)}_{\alpha\beta}$ denotes the $(\alpha,\beta)$ element of the block associated with sites $i$ and $j$ in the covariance matrix, $\sigma^{\alpha}_{i}$ ($\alpha=x,y,z$) are Pauli matrices, and $\langle \cdots\rangle$ represents the expectation values for a quantum state. It can be seen that the diagonal blocks ($i=j$) correspond to local spin fluctuations, which are the characteristics of traditional symmetry-breaking phase transitions due to the presence of local order parameters. On the other hand, the off-diagonal blocks ($i\neq j$) include all the non-local fluctuations between different spin components throughout the system, which may involve subtle signatures of topological phase transitions. 

As a consequence, it is expected that PCA can give a comprehensive statistical characterization of both conventional symmetry-breaking and topological quantum phase transitions by capturing both local and non-local fluctuation patterns in the data. Below, we will present a concrete example for demonstration.

\subsection{Example: 1D transverse-field Ising}
\label{sec: 1dtfim}

\begin{figure}[ht]
    \centering
    \includegraphics[width=0.48\textwidth]{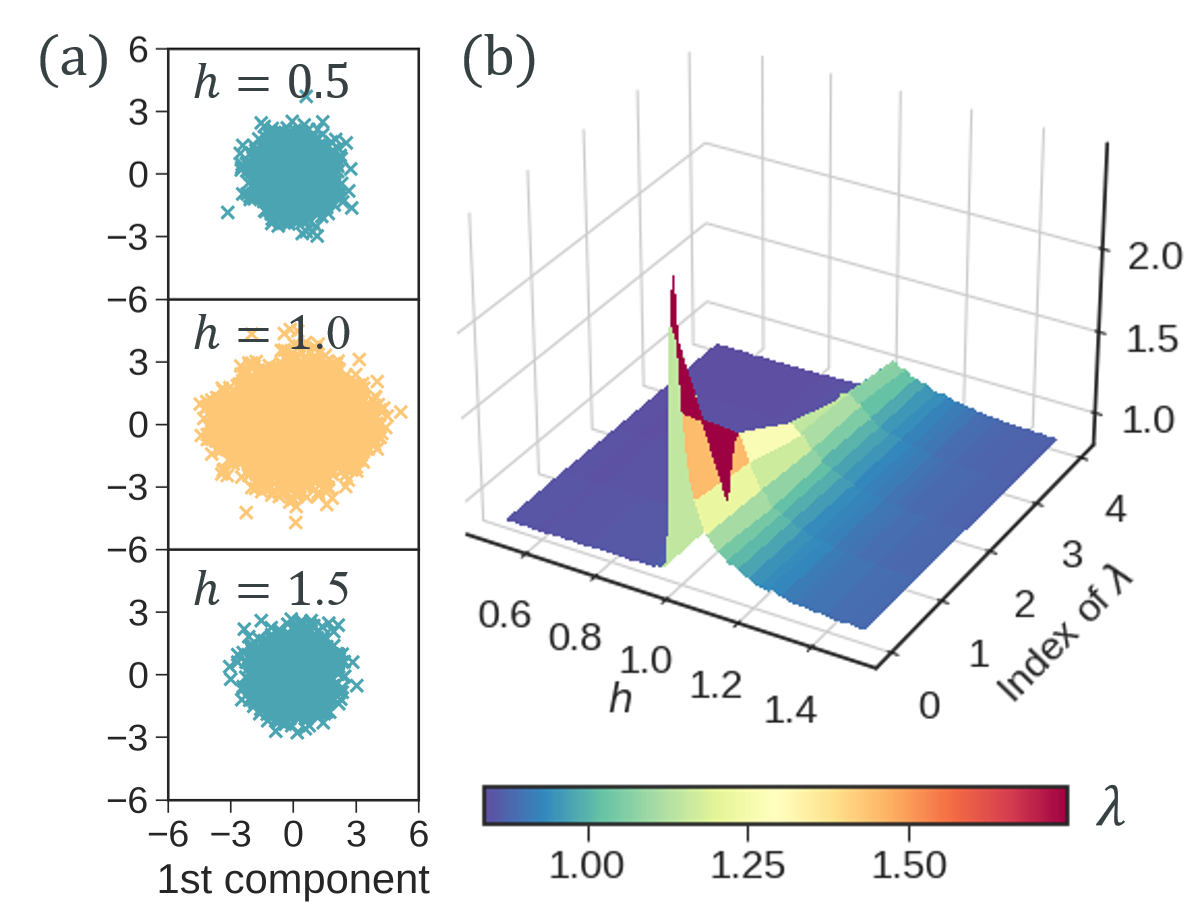}
    \caption{
    PCA results for the 1D transverse-field Ising model with $L=200$. 
    (a) The distribution of spin configurations projected onto the first and second principal components for $h = 0.5,\ 1.0,\ 1.5$.
    (b) The standard deviations of the first few principal components as a function of $h$, showing clear peaks near the critical point $h=1$. 
    } 
   \label{fig: 1dtfim}
\end{figure}

Here we take the 1D transverse-field Ising model (TFIM) as an illustrative example, which is a paradigm for studying quantum phase transitions: 
\begin{equation}
    H_{\text{TFIM}} = 
    -\sum_{i=1}^{L-1}\sigma^{z}_{i}\sigma^{z}_{i+1}
    -h\sum_{i=1}^{L}\sigma^{x}_{i},
    \label{eqn: 1dtfim}
\end{equation}
where $h$ denotes the external field. At zero temperature, this model has a symmetry-breaking quantum phase transition at the critical field strength $h=1$, separating a ferromagnetic phase ($h<1$) from a paramagnetic phase $(h>1)$. Here, we use the density matrix renormalization group (DMRG) algorithm to compute the ground states for system sizes $L=200$ and $h$ ranging from $0.5$ to $1.5$.

Fig.~\ref{fig: 1dtfim}(a) shows the distribution of spin configurations projected onto the first and second principal components. When deep in the parameter regions of two phases ($h=0.5, 1.5$), the ground state has low quantum fluctuation, and thus the data distribution is mainly governed by the randomized choice of measured Pauli bases. This inherent fluctuation should be independent of the quantum phase and remain approximately constant across different parameter regimes. On the other hand, as the system approaches the critical point ($h=1.0$), the distribution of projected spin configurations becomes significantly broader, indicating that quantum fluctuation becomes dominant and substantially contributes to the overall data variance. 

To quantitatively characterize this behavior, in Fig.~\ref{fig: 1dtfim}(b), we present the standard deviations of the leading principal components (denoted by $\lambda_1$, $\lambda_2$, $\cdots$) as functions of $h$. One could observe clear peaks near the transition point, particularly in $\lambda_1$ and $\lambda_2$, reflecting the enhanced overall statistical fluctuation due to the competition between the two phases. These results suggest that the leading standard deviations of the principal components can serve as effective indicators of the underlying quantum phase transition. 

In the rest of this paper, we will present more examples encompassing both symmetry-breaking and topological quantum phase transitions. Furthermore, we analyze the distinctions between these two types of transitions and demonstrate how our PCA-based approach can effectively differentiate them. This capability distinguishes our method from conventional supervised and other unsupervised machine learning techniques.

\section{Identification of Topological Phase Transitions} 
\label{sec: identification-topological}
\subsection{1D XZX cluster-Ising}
\label{sec: cluster-ising}

\begin{figure}[ht]
    \centering
    \includegraphics[width=0.48\textwidth]{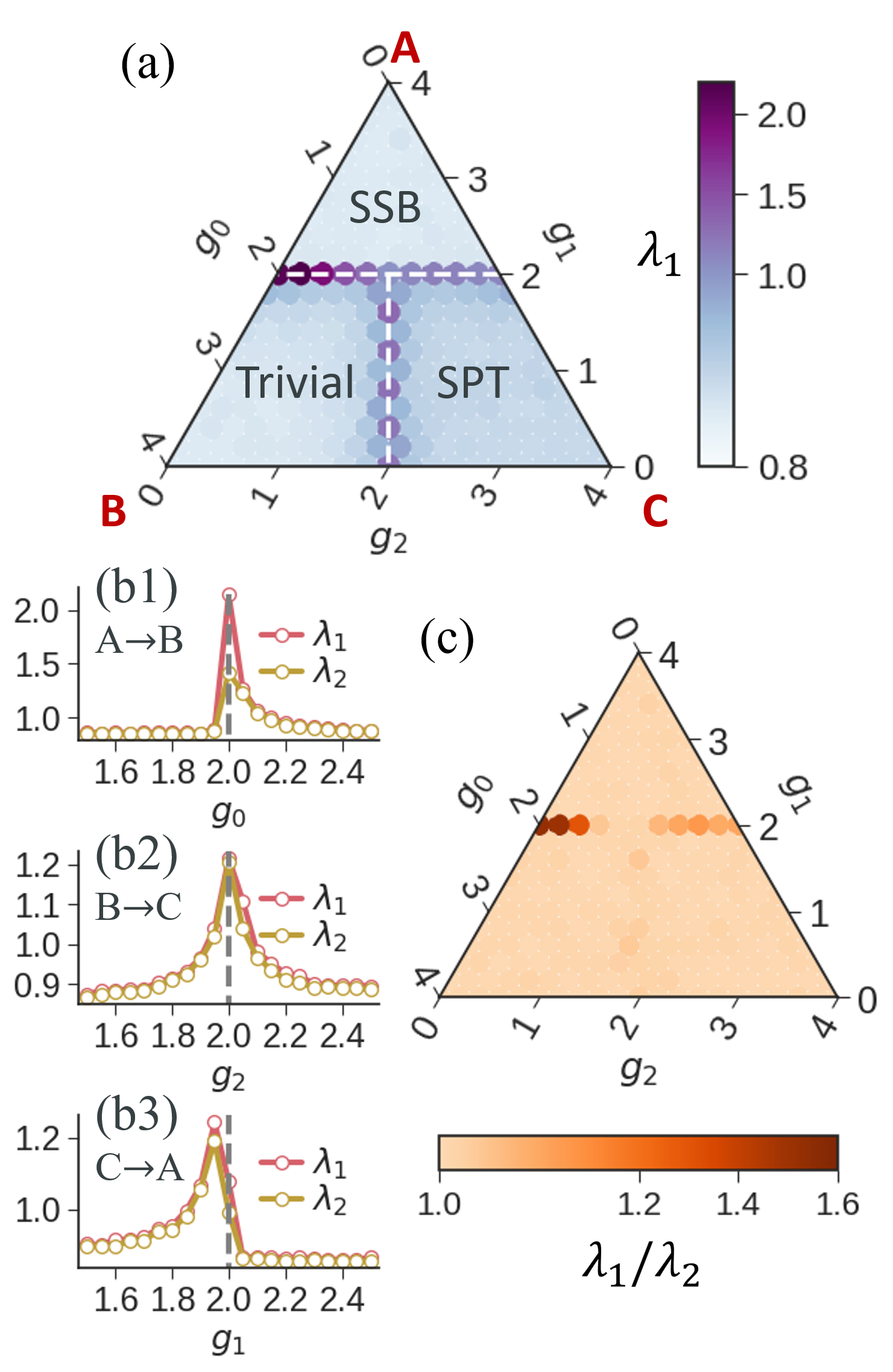}
    \caption{
    PCA results for the 1D XZX cluster-Ising model with $L=200$. 
    (a) The PCA $\lambda_1$ (colored dots) plotted over the ternary phase diagram under the constraint $g_0+g_1+g_2=4$. The dashed lines indicate the theoretical phase boundaries. 
    (b1-b3) The values of $\lambda_1$ and $\lambda_2$ along the three edges of the ternary phase diagram. The points A, B, and C mark the three corners of the diagram, as defined in panel (a).
    (c) The ratio $\lambda_1 / \lambda_2$ plotted over the ternary phase diagram.
    }
    \label{fig: cluster-ising}
\end{figure}

We continue to apply our approach to the 1D XZX cluster-Ising model \cite{paper_cluster-ising_1, paper_cluster-ising_2}, which extends the 1D TFIM by incorporating additional clustering interactions:
\begin{equation}
\begin{aligned}
    H_{\text{XZX-cluster}}= 
    &-g_{0}\sum_{i=1}^{L}\sigma^{z}_{i}
    -g_{1}\sum_{i=1}^{L-1}\sigma^{x}_{i}\sigma^{x}_{i+1} \\
  &-g_{2}\sum_{i=1}^{L-2}\sigma^{x}_{i} \sigma^{z}_{i+1} \sigma^{x}_{i+2},  
    \label{eqn: cluster-ising}
\end{aligned}
\end{equation}
where $g_0$ is the external field, $g_1$ represents the coupling strength, and $g_2$ controls the strength of the three-body XZX interaction. Fig.~\ref{fig: cluster-ising}(a) shows the ternary phase diagram under the constraint $g_0+g_1+g_2=4$. Along the left edge of the triangle ($g_2=0$), the model reduces to the previously discussed 1D TFIM, exhibiting a ferromagnetic phase with spontaneous symmetry breaking (SSB) and a trivial paramagnetic phase. By introducing the XZX term, the model supports an additional cluster phase that is associated with symmetry-protected topological (SPT) order. This enriches the phase diagram with both symmetry-breaking and topological transitions.

In Fig.~\ref{fig: cluster-ising}(a), we follow the same procedure as in the TFIM case and show the PCA $\lambda_1$ over the ternary phase diagram. We see that $\lambda_1$ reaches its maximum near all phase boundaries, including the SSB–SPT and SPT–trivial boundaries. In addition, we present how PCA $\lambda_1$ and $\lambda_2$ vary along with the three edges of the ternary phase diagram in Figs.~\ref{fig: cluster-ising}(b1)-\ref{fig: cluster-ising}(b3). These results confirm the effectiveness of our PCA approach in detecting both conventional symmetry-breaking and topological phase transitions.

However, one could observe a distinction in the relative magnitudes of $\lambda_1$ and $\lambda_2$ in Figs.~\ref{fig: cluster-ising}(b1)-\ref{fig: cluster-ising}(b3): $\lambda_1$ is significantly larger than $\lambda_2$ in the SSB-Trivial case, whereas the gap becomes less clear in the SSB-SPT and Trivial-SPT cases. To visualize this distinction more clearly, we plot the ratio of $\lambda_1$ to $\lambda_2$ over the ternary phase diagram in Fig.~\ref{fig: cluster-ising}(c). We observe that $\lambda_1/\lambda_2$ is typically $>1.5$ for SSB-trivial transitions, slightly greater than 1.0 for the SSB-SPT transitions, and about $1.0$ for SPT-Trivial transitions as well as within each phase. However, this distinction becomes less sharp near tricritical boundaries (i.e., center of the ternary phase diagram) due to the finite-size effects.

We note that these characteristic values of the $\lambda_1/\lambda_2$ ratio could be a qualitative diagnostic to distinguish the nature of phase transitions, especially those involving topological phase transitions. Besides, similar behaviors of the ratio $\lambda_1/\lambda_2$ are also observed in other models, as we demonstrate in the subsequent sections.

\subsection{1D bond-alternating XXZ}
\label{sec: alternating-xxz}
We now consider another 1D topological spin system: the 1D bond-alternating XXZ model \cite{paper_alternating-xxz_1}, which extends the conventional 1D XXZ model by introducing bond alternation:
\begin{equation}
    \begin{aligned}
        H_{\text{XXZ-alter}} = \sum_{i}
        &[1+(-1)^{i}\delta]
        (
        \sigma^{x}_{i}\sigma^{x}_{i+1} +
        \sigma^{y}_{i}\sigma^{y}_{i+1} \\
        &+\Delta\sigma^{z}_{i}\sigma^{z}_{i+1}
        ),
        \label{eqn: alternating-xxz}
    \end{aligned}
\end{equation}
where $\delta$ represents the bond alternation describing the dimerization induced by the Spin-Peierls instability \cite{book_giamarchi}, and $\Delta$ denotes the strength of the anisotropy. 

\begin{figure}[ht]
    \centering
    \includegraphics[width=0.48\textwidth]{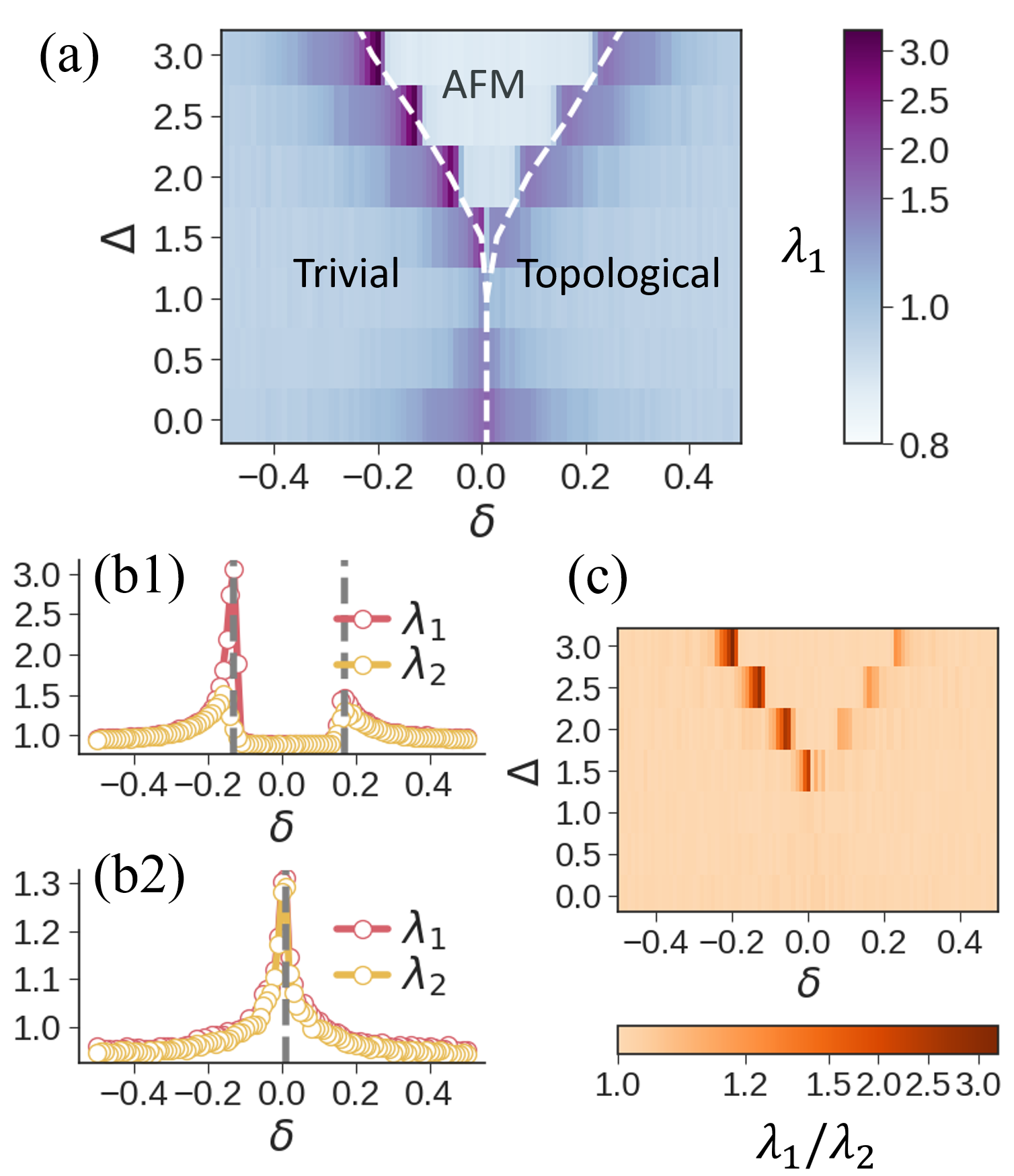}
    \caption{
    PCA results for the 1D bond-alternating XXZ chain with $L=200$. 
    (a) The PCA $\lambda_1$ plotted over the phase diagram. The white dashed lines indicate the phase boundary, determined by the peaks of entanglement entropy for $L$=200. 
    (b1-b2) The values of $\lambda_1$ and $\lambda_2$ plotted as functions of $\delta$ for fixed $\Delta=2.5$ and $\Delta=0.5$, respectively. 
    (c) The ratio of $\lambda_1$ to $\lambda_2$ plotted over the phase diagram.
    }
    \label{fig: alternating-xxz}
\end{figure}

In Fig.~\ref{fig: alternating-xxz}(a), we present the phase diagram of this model along with the PCA $\lambda_1$. In the absence of $\Delta$, the model is equivalent to the Su-Schrieffer-Heeger (SSH) chain \cite{paper_SSH}, whose ground-state exhibits classical dimer order (trivial phase) for $\delta<0$ and Haldane dimer order (topological phase) for $\delta>0$. As the anisotropy $\Delta$ increases sufficiently, a competing N{\'e}el order emerges, leading to a symmetry-breaking antiferromagnetic (AFM) phase. Notably, we see that the phase boundaries identified by our PCA $\lambda_1$ are very close to those obtained from DMRG calculations (white lines).

In Figs.~\ref{fig: alternating-xxz}(b1) and \ref{fig: alternating-xxz}(b2), we show the $\lambda_1$ and $\lambda_2$ as functions of $\delta$ for $\Delta=2.5$ and $\Delta=0.5$, respectively. Furthermore, in Fig.~\ref{fig: alternating-xxz}(c), we plot the ratio $\lambda_1/\lambda_2$ over the entire phase diagram to examine the distinctions between these different types of phase boundaries. As in the case of the 1D XZX cluster-Ising model, we can observe similar patterns: the ratio is typically $>1.5$ for AFM–Trivial transitions, around $<1.2$ for AFM-Topological transitions, and about $1.0$ for Trivial-Topological transitions. As a result, these distinct behaviors further support the capability of our PCA-based approach in identifying both the existence and nature of quantum phase transitions. 

We will further discuss this feature and its implications in Sec.~\ref{sec: discussion}.

\section{Application to Two-Dimensional Quantum Systems} 
\label{sec: 2d-quantum}
\subsection{2D transeverse-field Ising}
\label{sec: 2d-tfim}

\begin{figure}[ht]
    \centering
    \includegraphics[width=0.48\textwidth]{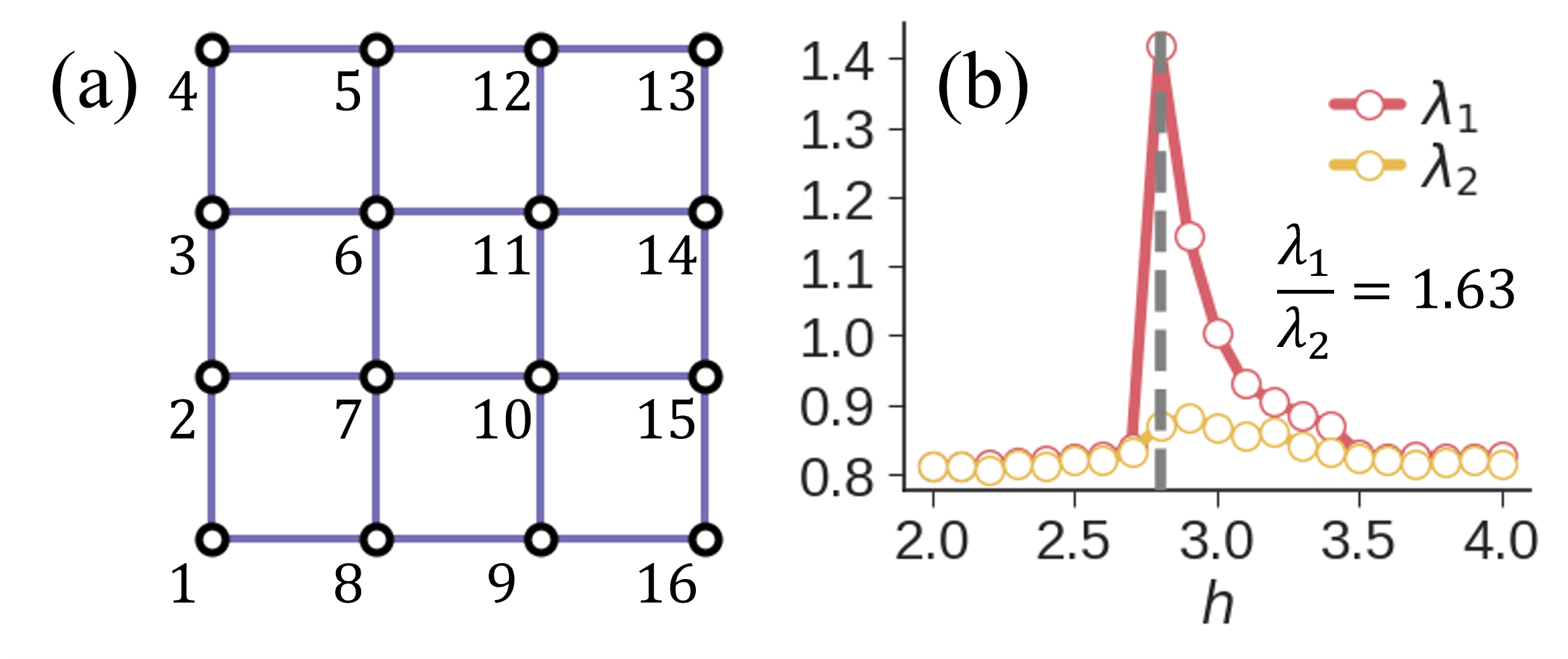}
    \caption{
    (a) Illustration of the 1D indexing scheme for a 2D lattice, using a $4 \times 4$ square lattice as an example. 
    (b) The PCA $\lambda_1$ and $\lambda_2$ for the 2D transverse-field Ising model on a $10 \times 10$ square lattice, plotted as functions of $h$. The notable peaks are observed near the critical point ($h_c\sim2.8$) with $\lambda_1/\lambda_2=1.63$. The dashed line denotes the phase boundary, determined by the peak of entanglement entropy.  
    }
    \label{fig: 2d-tfim}
\end{figure}

It is of interest to examine the applicability of our approach for two-dimensional quantum systems. As a representative example, we consider the 2D TFIM on a square lattice \cite{paper_2d-tfim}:
\begin{equation}
    H_{\text{2D-TFIM}} = 
    -\sum_{\langle i,j\rangle}\sigma^{z}_{i}\sigma^{z}_{j}
    -h\sum_{i}\sigma^{x}_{i}
    \label{eqn: 2d-tfim}
\end{equation}
where $\langle i,j \rangle$ denotes nearest-neighbor pairs on a 2D square lattice, and $h$ is the strength of the transverse magnetic field. To match the input format required by PCA (namely, a one-dimensional vector representation), we relabel each site of the 2D square lattice using a 1D indexing scheme, as illustrated in Fig.~\ref{fig: 2d-tfim}(a). As a result, a spin configuration on a $L\times L$ square lattice is represented as a vector of length $3L^2$, with each site encoded in a 3-dimensional Euclidean form.

In Fig.~\ref{fig: 2d-tfim}(b), we consider a $10\times10$ square lattice and show the PCA $\lambda_1$ and $\lambda_2$ as functions of $h$. The emergence of peaks with a substantial gap between $\lambda_1$ and $\lambda_2$ ($\lambda_1/\lambda_2 = 1.63$) is observed near the critical point ($h_c\sim2.8$). This characteristic behavior signals the existence of a symmetry-breaking quantum phase transition between ferromagnetic and paramagnetic phases, which is consistent with our theoretical expectations. Therefore, these results demonstrate the applicability of our PCA-based approach to identifying quantum phase transitions in two-dimensional systems.

\subsection{2D Kitaev honeycomb}
\label{sec: 2d-kitaev}
To further examine a topological phase transition in a 2D quantum system, we consider the 2D Kitaev model on the honeycomb lattice \cite{paper_2d-kitaev}:
\begin{equation}
    \begin{aligned}
        H_{\text{2D-Kitaev}} = 
        &-J_x\sum_{\langle i,j\rangle_x}\sigma^{x}_{i}\sigma^{x}_{j}
        -J_y\sum_{\langle i,j\rangle_y}\sigma^{y}_{i}\sigma^{y}_{j} \\
        &-J_z\sum_{\langle i,j\rangle_z}\sigma^{z}_{i}\sigma^{z}_{j},
    \end{aligned}
    \label{eqn: 2d-kitaev}
\end{equation}
where $\langle i,j\rangle_\alpha$ ($\alpha=x,y,z$) denotes nearest-neighbor pairs connected by $\alpha$-type bonds on the lattice, and $J_x$, $J_y$, $J_z$ are the anisotropic coupling strengths. This model is exactly solvable and is known to host a quantum spin liquid (QSL) ground state.

\begin{figure}[ht]
    \centering
    \includegraphics[width=0.48\textwidth]{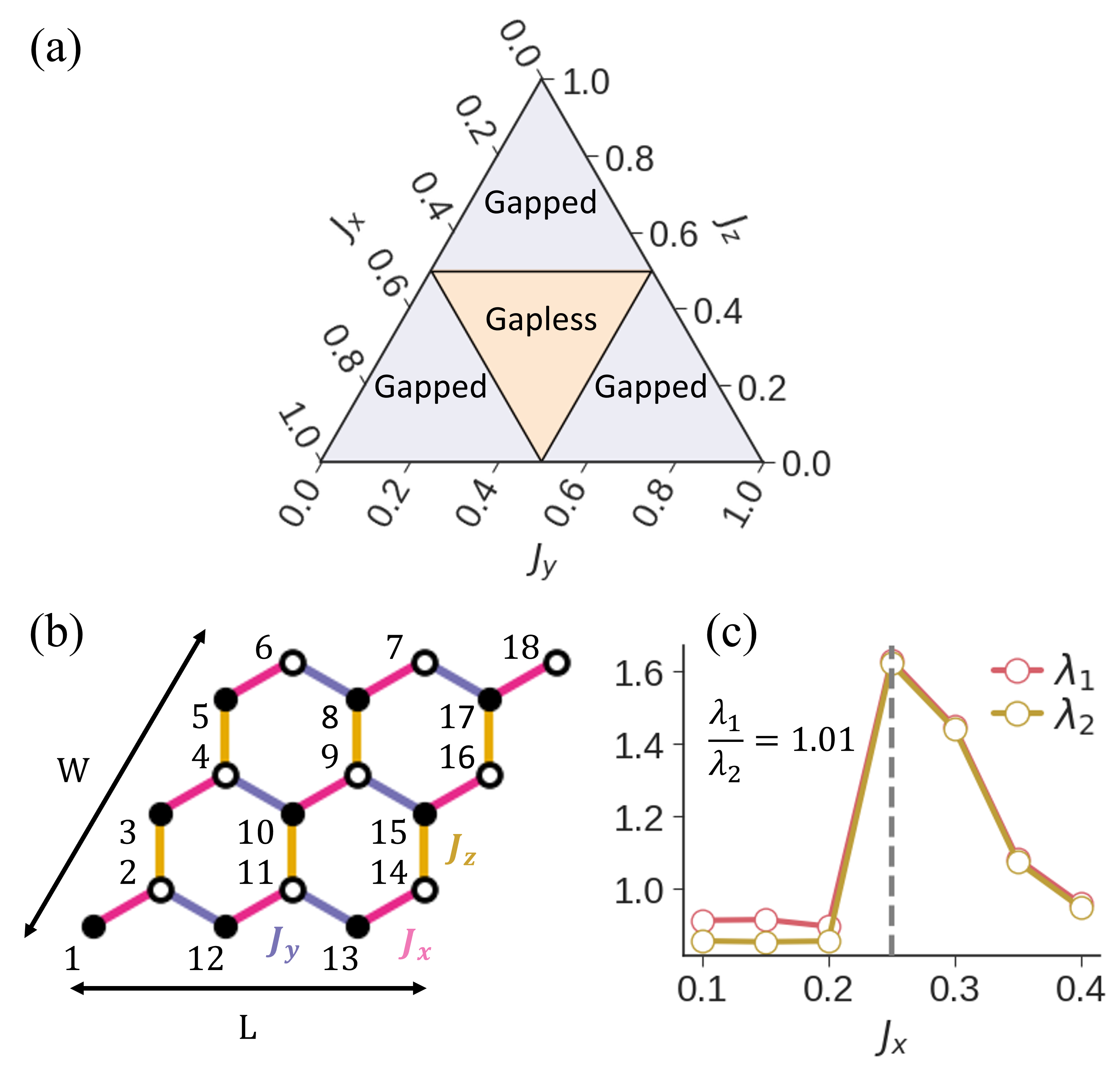}
    \caption{ 
    (a) The ternary phase diagram of 2D Kitaev honeycomb model under the constraint $J_x+J_y+J_z=1$.
    (b) Illustration of the 1D indexing scheme for a 2D $L \times W$ honeycomb lattice, using $L$=3 and $W$=3 as an example.
    (c) The PCA $\lambda_1$ and $\lambda_2$ for the 2D Kitaev model along the parameter path $J_x=J_y$. Here we consider a honeycomb lattice with $L$=8 and $W$=8. The dashed line denotes the topological phase boundary $J_x=J_y=0.25$.
    }
    \label{fig: 2d-kitaev}
\end{figure}

In Fig.~\ref{fig: 2d-kitaev}(a), we show the ternary phase diagram under the constraint $J_x+J_y+J_z=1$. When one of the couplings dominates over the others, the system is in a gapped quantum spin liquid phase characterized by localized Majorana fermions and non-Abelian anyonic excitations. In contrast, when the couplings are more balanced, the system enters a gapless quantum spin liquid phase, in which  Majorana fermions give rise to a Dirac-like spectrum. The transition between the gapless and gapped phases does not involve any symmetry breaking and is thus a topological phase transition.

In Figs.~\ref{fig: 2d-kitaev}(b) and \ref{fig: 2d-kitaev}(c), we show the 1D indexing scheme for a 2D honeycomb lattice and present the PCA $\lambda_1$ and $\lambda_2$ along the parameter path $J_x=J_y$. It can be observed that $\lambda_1$ and $\lambda_2$ exhibit similar peak structures near $J_x=J_y=0.25$ with $\lambda_1/\lambda_2 = 1.01$, which is a clear signature of the topological phase transition. This behavior is in agreement with the known transition between the gapless and gapped QSL regimes, thereby showing the capability of $\lambda_1/\lambda_2$ to distinguish topological phase transitions in 2D quantum systems.

\section{Discussion}
\label{sec: discussion} 
\subsection{PCA fluctuation structures across symmetry-breaking and topological transitions}

\begin{figure}[ht]
    \centering
    \includegraphics[width=0.48\textwidth]{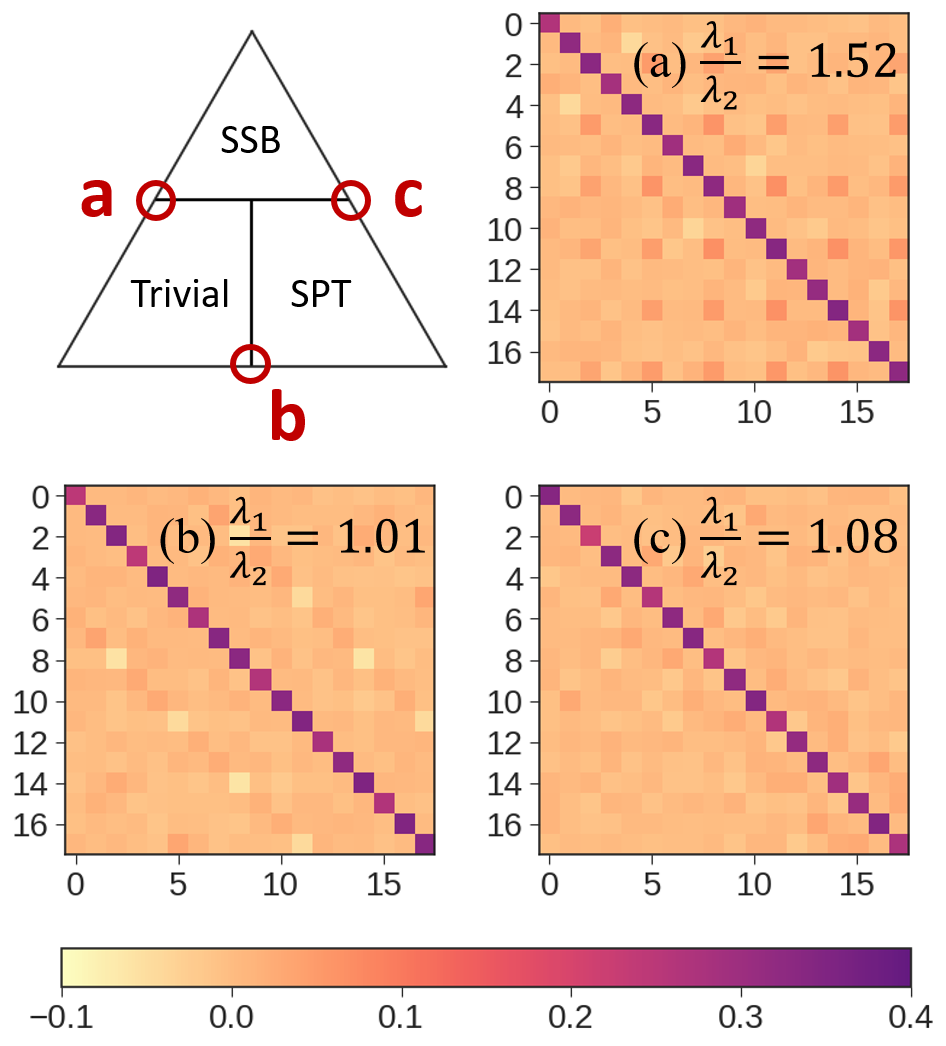}
    \caption{
    PCA covariance matrices for the 1D XZX cluster-Ising model with $L$=200. Here we only display the upper-left 18 $\times$ 18 block for illustration, with value of $\lambda_1/\lambda_2$ obtained using $N$=2000 \textit{in-situ} measurement data. 
    (a), (b) and (c) denote the SSB-Trivial, Topological-Trivial, and SSB-Topological boundaries, respectively, as shown in the upper-left panel. 
    }
    \label{fig: discussion-cluster}
\end{figure}

In this work, we have demonstrated that the ratio $\lambda_1/\lambda_2$ could be a useful signature for distinguishing symmetry-breaking and topological phase transitions in both 1D and 2D quantum systems. As a result, it is instructive to investigate why the structure of statistical fluctuation extracted by PCA connects to the nature of quantum fluctuation. To this end, we consider the 1D XZX cluster-Ising model as a benchmark and present the PCA covariance matrices near different phase boundaries in Fig.~\ref{fig: discussion-cluster}. 

We note that $\lambda_1$ and $\lambda_2$ can also be directly computed by using Eq.~(\ref{eqn: block}) and the DMRG algorithm. The corresponding values of $\lambda_1/\lambda_2$ in Figs.~\ref{fig: discussion-cluster}(a)-(c) are 1.57, 1.00, and 1.26, respectively, which are close to those obtained using \textit{in-situ} measurement data (1.52, 1.01 and 1.08). Therefore, our PCA-based framework could offer a robustness analysis with interpretable perspectives on the underlying physical mechanisms behind different phase transitions.

In particular, the symmetry-breaking case (panel (a)) has obvious long-range fluctuations. This is significantly different from other topological phase transitions (panels (b) and (c)). The underlying physics could be qualitatively understood as follows: for the symmetry-breaking transitions, the ground states on either side of the transition are macroscopically distinct, and often related to the emergence of long-range order. Therefore, it is expected that the competition between phases gives rise to certain long-range fluctuations throughout the system, as reflected in the off-diagonal components of the covariance matrix in Fig.~\ref{fig: discussion-cluster}(a). This leads to a strong directional variance in the dataset, thus resulting in a clear separation between $\lambda_1$ and $\lambda_2$. In contrast, topological phase transitions do not involve local order parameters or symmetry breaking, and therefore lack conventional long-range fluctuations, as shown in Figs.~\ref{fig: discussion-cluster}(b). The ground states on both sides of the transition are locally similar, with subtle distinctions emerging only through non-local or entanglement-related features. As a result, the statistical fluctuations in the dataset are more isotropic, making a smaller value of the PCA ratio, with $\lambda_1/\lambda_2 \approx 1$. However, for the SSB-Topological boundary (see Fig.~\ref{fig: discussion-cluster}(c)), while one side of the transition supports long-range order, the other is characterized by topological correlations. This leads to intermediate patterns in the covariance structure, where directional variance is present but less significant compared to pure symmetry-breaking transitions. Consequently, the $\lambda_1/\lambda_2$ would take moderate values between the two extremes.

\subsection{Comparison between different machine learning approaches}

\begin{table*}[ht]
    \centering
    \renewcommand{\arraystretch}{2.0} 
    \tabcolsep=1pt 
    \begin{tabular}{c c c c c}
    \hline\hline
    \textbf{Aspects} 
    & Supervised 
    & Conventional Unsupervised
    & Self-Supervised 
    & *Other Unsupervised \\
    \hline
    \textbf{Typical Task} 
    & Classify phases 
    & Discover clusters 
    & Simulate state function
    & Analyze fluctuation \\
    \textbf{Label Requirement} 
    & Yes 
    & No 
    & No 
    & No \\
    \textbf{Applicability to New Systems} 
    & No 
    & Yes 
    & Yes 
    & Yes \\
    \textbf{Classification of Transitions} 
    & No 
    & No
    & Yes 
    & Depends on Method \\
    \hline\hline
    \end{tabular}
    \caption{
    Comparison between supervised, conventional unsupervised, self-supervised, and other unsupervised learning approaches in the context of identifying many-body phase transitions. Our PCA-based framework belongs to the category of other unsupervised learning, with the capability to not only identify phase boundaries but also classify the nature of the transition. Such dual capability has been studied only in a few advanced unsupervised approaches \cite{paper_USL_14, paper_USL_16}
    }
    \label{tab: ML-comparison}
\end{table*}

To clarify the distinctions among previous machine learning approaches for phase transition studies, a concise comparison is provided in Table~\ref{tab: ML-comparison}. Supervised learning methods rely on labeled data to classify phases, making them highly effective once phase information or order parameters are known. However, this dependence also limits their applicability to exploring new or poorly understood systems, where such labels are not available. By contrast, conventional unsupervised learning techniques cluster raw measurement data without prior labels, and therefore can in principle be applied to unfamiliar systems. Nevertheless, their outcomes often lack clear physical interpretation and may not reliably distinguish different types of phase transitions. In this context, self-supervised learning provides a potential alternative by simulating the system state function from \textit{in-situ} spin configurations measured on a specific known basis. While practical when external experimental control is well characterized, this requirement can be restrictive. 

On the other hand, our PCA-based framework falls into the category of other unsupervised learning, which typically extracts information from intrinsic data fluctuations. In particular, unlike self-supervised methods, our approach does not require such assumptions and instead incorporates spin configurations measured on randomized bases. This not only broadens its applicability to systems without well-defined external controls but also offers a general and physically interpretable route for analyzing quantum phase transitions, making it particularly suitable for exploring unknown quantum many-body systems.

\section{Summary}
\label{sec: summary}
In summary, we have presented an integrated methodology for detecting quantum phase transitions by applying unsupervised principal component analysis (PCA) to randomized measurement data. By encoding spin configurations obtained from randomized Pauli measurements and analyzing their statistical fluctuations, we demonstrate that the PCA $\lambda_1$ serves as a sensitive indicator of criticality. Our numerical studies on a variety of both 1D and 2D quantum systems confirm the effectiveness and generality of this approach. Furthermore, the observed fluctuation structure, particularly the ratio $\lambda_1/\lambda_2$, offers a qualitative classification of different types of phase transitions. 

As a result, these results highlight the potential of combining randomized quantum measurements with unsupervised learning techniques to uncover complex many-body phenomena, even in the absence of explicit knowledge of the Hamiltonian. Future directions include extending this framework to finite-temperature systems, generalizing to fermionic or bosonic systems, or integrating with more advanced machine learning methods to enable deeper insights into quantum matter.

\section{Acknowledgement} 
\label{sec: acknowledgement}
The DMRG simulation is implemented by the ITensors.jl package \cite{ITensor, ITensor-r0.3}. We thank Hsiu-Chuan Hsu, Jhih-Shih You, Yen-Hsiang Lin, and Po-Chung Chen for the valuable discussion. This work is supported by the National Center for Theoretical Sciences, the Higher Education Sprout Project funded by the Ministry of Science and Technology, and the Ministry of Education in Taiwan. DWW is supported under the grant MOST 110-2112-M-007-036-MY3.

\bibliography{main}

\providecommand{\noopsort}[1]{}\providecommand{\singleletter}[1]{#1}%
\begin{thebibliography}{45}%
\makeatletter
\providecommand \@ifxundefined [1]{%
 \@ifx{#1\undefined}
}%
\providecommand \@ifnum [1]{%
 \ifnum #1\expandafter \@firstoftwo
 \else \expandafter \@secondoftwo
 \fi
}%
\providecommand \@ifx [1]{%
 \ifx #1\expandafter \@firstoftwo
 \else \expandafter \@secondoftwo
 \fi
}%
\providecommand \natexlab [1]{#1}%
\providecommand \enquote  [1]{``#1''}%
\providecommand \bibnamefont  [1]{#1}%
\providecommand \bibfnamefont [1]{#1}%
\providecommand \citenamefont [1]{#1}%
\providecommand \href@noop [0]{\@secondoftwo}%
\providecommand \href [0]{\begingroup \@sanitize@url \@href}%
\providecommand \@href[1]{\@@startlink{#1}\@@href}%
\providecommand \@@href[1]{\endgroup#1\@@endlink}%
\providecommand \@sanitize@url [0]{\catcode `\\12\catcode `\$12\catcode `\&12\catcode `\#12\catcode `\^12\catcode `\_12\catcode `\%12\relax}%
\providecommand \@@startlink[1]{}%
\providecommand \@@endlink[0]{}%
\providecommand \url  [0]{\begingroup\@sanitize@url \@url }%
\providecommand \@url [1]{\endgroup\@href {#1}{\urlprefix }}%
\providecommand \urlprefix  [0]{URL }%
\providecommand \Eprint [0]{\href }%
\providecommand \doibase [0]{https://doi.org/}%
\providecommand \selectlanguage [0]{\@gobble}%
\providecommand \bibinfo  [0]{\@secondoftwo}%
\providecommand \bibfield  [0]{\@secondoftwo}%
\providecommand \translation [1]{[#1]}%
\providecommand \BibitemOpen [0]{}%
\providecommand \bibitemStop [0]{}%
\providecommand \bibitemNoStop [0]{.\EOS\space}%
\providecommand \EOS [0]{\spacefactor3000\relax}%
\providecommand \BibitemShut  [1]{\csname bibitem#1\endcsname}%
\let\auto@bib@innerbib\@empty
\bibitem [{\citenamefont {Savary}\ and\ \citenamefont {Balents}(2016)}]{intro_QSL_1}%
  \BibitemOpen
  \bibfield  {author} {\bibinfo {author} {\bibfnamefont {L.}~\bibnamefont {Savary}}\ and\ \bibinfo {author} {\bibfnamefont {L.}~\bibnamefont {Balents}},\ }\bibfield  {title} {\bibinfo {title} {Quantum spin liquids: a review},\ }\href {https://doi.org/10.1088/0034-4885/80/1/016502} {\bibfield  {journal} {\bibinfo  {journal} {Rep. Prog. Phys}\ }\textbf {\bibinfo {volume} {80}},\ \bibinfo {pages} {016502} (\bibinfo {year} {2016})}\BibitemShut {NoStop}%
\bibitem [{\citenamefont {Zhou}\ \emph {et~al.}(2017)\citenamefont {Zhou}, \citenamefont {Kanoda},\ and\ \citenamefont {Ng}}]{intro_QSL_2}%
  \BibitemOpen
  \bibfield  {author} {\bibinfo {author} {\bibfnamefont {Y.}~\bibnamefont {Zhou}}, \bibinfo {author} {\bibfnamefont {K.}~\bibnamefont {Kanoda}},\ and\ \bibinfo {author} {\bibfnamefont {T.-K.}\ \bibnamefont {Ng}},\ }\bibfield  {title} {\bibinfo {title} {Quantum spin liquid states},\ }\href {https://doi.org/10.1103/RevModPhys.89.025003} {\bibfield  {journal} {\bibinfo  {journal} {Rev. Mod. Phys.}\ }\textbf {\bibinfo {volume} {89}},\ \bibinfo {pages} {025003} (\bibinfo {year} {2017})}\BibitemShut {NoStop}%
\bibitem [{\citenamefont {Bernevig}(2013)}]{book_Bernevig}%
  \BibitemOpen
  \bibfield  {author} {\bibinfo {author} {\bibfnamefont {B.~A.}\ \bibnamefont {Bernevig}},\ }\href {https://doi.org/doi:10.1515/9781400846733} {\emph {\bibinfo {title} {Topological Insulators and Topological Superconductors}}}\ (\bibinfo  {publisher} {Princeton University Press},\ \bibinfo {address} {Princeton},\ \bibinfo {year} {2013})\BibitemShut {NoStop}%
\bibitem [{\citenamefont {Vermersch}\ \emph {et~al.}(2019)\citenamefont {Vermersch}, \citenamefont {Elben}, \citenamefont {Sieberer}, \citenamefont {Yao},\ and\ \citenamefont {Zoller}}]{intro_randomized-measurement_1}%
  \BibitemOpen
  \bibfield  {author} {\bibinfo {author} {\bibfnamefont {B.}~\bibnamefont {Vermersch}}, \bibinfo {author} {\bibfnamefont {A.}~\bibnamefont {Elben}}, \bibinfo {author} {\bibfnamefont {L.~M.}\ \bibnamefont {Sieberer}}, \bibinfo {author} {\bibfnamefont {N.~Y.}\ \bibnamefont {Yao}},\ and\ \bibinfo {author} {\bibfnamefont {P.}~\bibnamefont {Zoller}},\ }\bibfield  {title} {\bibinfo {title} {Probing scrambling using statistical correlations between randomized measurements},\ }\href {https://doi.org/10.1103/PhysRevX.9.021061} {\bibfield  {journal} {\bibinfo  {journal} {Phys. Rev. X}\ }\textbf {\bibinfo {volume} {9}},\ \bibinfo {pages} {021061} (\bibinfo {year} {2019})}\BibitemShut {NoStop}%
\bibitem [{\citenamefont {et~al.}(2020{\natexlab{a}})}]{intro_randomized-measurement_2}%
  \BibitemOpen
  \bibfield  {author} {\bibinfo {author} {\bibfnamefont {T.~B.}\ \bibnamefont {et~al.}},\ }\bibfield  {title} {\bibinfo {title} {Probing rényi entanglement entropy via randomized measurements},\ }\href {https://doi.org/10.1126/sciadv.aaz3666} {\bibfield  {journal} {\bibinfo  {journal} {Science}\ }\textbf {\bibinfo {volume} {6}},\ \bibinfo {pages} {eaaz3666} (\bibinfo {year} {2020}{\natexlab{a}})}\BibitemShut {NoStop}%
\bibitem [{\citenamefont {et~al.}(2020{\natexlab{b}})}]{intro_randomized-measurement_3}%
  \BibitemOpen
  \bibfield  {author} {\bibinfo {author} {\bibfnamefont {A.~E.}\ \bibnamefont {et~al.}},\ }\bibfield  {title} {\bibinfo {title} {Many-body topological invariants from randomized measurements in synthetic quantum matter},\ }\href {https://doi.org/10.1126/sciadv.aaz3666} {\bibfield  {journal} {\bibinfo  {journal} {Sci. Adv.}\ }\textbf {\bibinfo {volume} {6}},\ \bibinfo {pages} {eaaz3666} (\bibinfo {year} {2020}{\natexlab{b}})}\BibitemShut {NoStop}%
\bibitem [{\citenamefont {Aaronson}(2018)}]{intro_classical-shadow_1}%
  \BibitemOpen
  \bibfield  {author} {\bibinfo {author} {\bibfnamefont {S.}~\bibnamefont {Aaronson}},\ }\bibfield  {title} {\bibinfo {title} {Shadow tomography of quantum states},\ }in\ \href@noop {} {\emph {\bibinfo {booktitle} {Proceedings of the 50th annual ACM SIGACT symposium on theory of computing}}}\ (\bibinfo {year} {2018})\ pp.\ \bibinfo {pages} {325--338}\BibitemShut {NoStop}%
\bibitem [{\citenamefont {Huang}\ \emph {et~al.}(2020)\citenamefont {Huang}, \citenamefont {Kueng},\ and\ \citenamefont {Preskill}}]{intro_classical-shadow_2}%
  \BibitemOpen
  \bibfield  {author} {\bibinfo {author} {\bibfnamefont {H.-Y.}\ \bibnamefont {Huang}}, \bibinfo {author} {\bibfnamefont {R.}~\bibnamefont {Kueng}},\ and\ \bibinfo {author} {\bibfnamefont {J.}~\bibnamefont {Preskill}},\ }\bibfield  {title} {\bibinfo {title} {Predicting many properties of a quantum system from very few measurements},\ }\href {https://doi.org/10.1038/s41567-020-0932-7} {\bibfield  {journal} {\bibinfo  {journal} {Nat. Phys.}\ }\textbf {\bibinfo {volume} {16}},\ \bibinfo {pages} {1050–1057} (\bibinfo {year} {2020})}\BibitemShut {NoStop}%
\bibitem [{\citenamefont {Carrasquilla}\ and\ \citenamefont {Melko}(2017)}]{paper_SL_1}%
  \BibitemOpen
  \bibfield  {author} {\bibinfo {author} {\bibfnamefont {J.}~\bibnamefont {Carrasquilla}}\ and\ \bibinfo {author} {\bibfnamefont {R.~G.}\ \bibnamefont {Melko}},\ }\bibfield  {title} {\bibinfo {title} {Machine learning phases of matter},\ }\href {https://doi.org/10.1038/nphys4035} {\bibfield  {journal} {\bibinfo  {journal} {Nat. Phys.}\ }\textbf {\bibinfo {volume} {13}},\ \bibinfo {pages} {431} (\bibinfo {year} {2017})}\BibitemShut {NoStop}%
\bibitem [{\citenamefont {Zhang}\ \emph {et~al.}(2019)\citenamefont {Zhang}, \citenamefont {Liu},\ and\ \citenamefont {Wei}}]{paper_SL_2}%
  \BibitemOpen
  \bibfield  {author} {\bibinfo {author} {\bibfnamefont {W.}~\bibnamefont {Zhang}}, \bibinfo {author} {\bibfnamefont {J.}~\bibnamefont {Liu}},\ and\ \bibinfo {author} {\bibfnamefont {T.-C.}\ \bibnamefont {Wei}},\ }\bibfield  {title} {\bibinfo {title} {Machine learning of phase transitions in the percolation and $xy$ models},\ }\href {https://doi.org/10.1103/PhysRevE.99.032142} {\bibfield  {journal} {\bibinfo  {journal} {Phys. Rev. E}\ }\textbf {\bibinfo {volume} {99}},\ \bibinfo {pages} {032142} (\bibinfo {year} {2019})}\BibitemShut {NoStop}%
\bibitem [{\citenamefont {Giannetti}\ \emph {et~al.}(2019)\citenamefont {Giannetti}, \citenamefont {Lucini},\ and\ \citenamefont {Vadacchino}}]{paper_SL_3}%
  \BibitemOpen
  \bibfield  {author} {\bibinfo {author} {\bibfnamefont {C.}~\bibnamefont {Giannetti}}, \bibinfo {author} {\bibfnamefont {B.}~\bibnamefont {Lucini}},\ and\ \bibinfo {author} {\bibfnamefont {D.}~\bibnamefont {Vadacchino}},\ }\bibfield  {title} {\bibinfo {title} {Machine learning as a universal tool for quantitative investigations of phase transitions},\ }\href {https://doi.org/https://doi.org10.1016/j.nuclphysb.2019.114639} {\bibfield  {journal} {\bibinfo  {journal} {Nucl. Phys. B}\ }\textbf {\bibinfo {volume} {944}},\ \bibinfo {pages} {114639} (\bibinfo {year} {2019})}\BibitemShut {NoStop}%
\bibitem [{\citenamefont {Shiina}\ \emph {et~al.}(2020)\citenamefont {Shiina}, \citenamefont {Mori}, \citenamefont {Okabe},\ and\ \citenamefont {Lee}}]{paper_SL_4}%
  \BibitemOpen
  \bibfield  {author} {\bibinfo {author} {\bibfnamefont {K.}~\bibnamefont {Shiina}}, \bibinfo {author} {\bibfnamefont {H.}~\bibnamefont {Mori}}, \bibinfo {author} {\bibfnamefont {Y.}~\bibnamefont {Okabe}},\ and\ \bibinfo {author} {\bibfnamefont {H.~K.}\ \bibnamefont {Lee}},\ }\bibfield  {title} {\bibinfo {title} {Machine-learning studies on spin models},\ }\href {https://doi.org/10.1038/s41598-020-58263-5} {\bibfield  {journal} {\bibinfo  {journal} {Sci Rep}\ }\textbf {\bibinfo {volume} {10}},\ \bibinfo {pages} {2177} (\bibinfo {year} {2020})}\BibitemShut {NoStop}%
\bibitem [{\citenamefont {van Nieuwenburg}\ \emph {et~al.}(2017)\citenamefont {van Nieuwenburg}, \citenamefont {Liu},\ and\ \citenamefont {Huber}}]{paper_SL_5}%
  \BibitemOpen
  \bibfield  {author} {\bibinfo {author} {\bibfnamefont {E.~P.~L.}\ \bibnamefont {van Nieuwenburg}}, \bibinfo {author} {\bibfnamefont {Y.-H.}\ \bibnamefont {Liu}},\ and\ \bibinfo {author} {\bibfnamefont {S.~D.}\ \bibnamefont {Huber}},\ }\bibfield  {title} {\bibinfo {title} {Learning phase transitions by confusion},\ }\href {https://doi.org/10.1038/nphys4037} {\bibfield  {journal} {\bibinfo  {journal} {Nat. Phys.}\ }\textbf {\bibinfo {volume} {13}},\ \bibinfo {pages} {435} (\bibinfo {year} {2017})}\BibitemShut {NoStop}%
\bibitem [{\citenamefont {Ch'ng}\ \emph {et~al.}(2017)\citenamefont {Ch'ng}, \citenamefont {Carrasquilla}, \citenamefont {Melko},\ and\ \citenamefont {Khatami}}]{paper_SL_6}%
  \BibitemOpen
  \bibfield  {author} {\bibinfo {author} {\bibfnamefont {K.}~\bibnamefont {Ch'ng}}, \bibinfo {author} {\bibfnamefont {J.}~\bibnamefont {Carrasquilla}}, \bibinfo {author} {\bibfnamefont {R.~G.}\ \bibnamefont {Melko}},\ and\ \bibinfo {author} {\bibfnamefont {E.}~\bibnamefont {Khatami}},\ }\bibfield  {title} {\bibinfo {title} {Machine learning phases of strongly correlated fermions},\ }\href {https://doi.org/10.1103/PhysRevX.7.031038} {\bibfield  {journal} {\bibinfo  {journal} {Phys. Rev. X}\ }\textbf {\bibinfo {volume} {7}},\ \bibinfo {pages} {031038} (\bibinfo {year} {2017})}\BibitemShut {NoStop}%
\bibitem [{\citenamefont {Rem}\ \emph {et~al.}(2019)\citenamefont {Rem}, \citenamefont {K{\"a}ming}, \citenamefont {Tarnowski}, \citenamefont {Asteria}, \citenamefont {Fl{\"a}schner}, \citenamefont {Becker}, \citenamefont {Sengstock},\ and\ \citenamefont {Weitenberg}}]{paper_SL_7}%
  \BibitemOpen
  \bibfield  {author} {\bibinfo {author} {\bibfnamefont {B.~S.}\ \bibnamefont {Rem}}, \bibinfo {author} {\bibfnamefont {N.}~\bibnamefont {K{\"a}ming}}, \bibinfo {author} {\bibfnamefont {M.}~\bibnamefont {Tarnowski}}, \bibinfo {author} {\bibfnamefont {L.}~\bibnamefont {Asteria}}, \bibinfo {author} {\bibfnamefont {N.}~\bibnamefont {Fl{\"a}schner}}, \bibinfo {author} {\bibfnamefont {C.}~\bibnamefont {Becker}}, \bibinfo {author} {\bibfnamefont {K.}~\bibnamefont {Sengstock}},\ and\ \bibinfo {author} {\bibfnamefont {C.}~\bibnamefont {Weitenberg}},\ }\bibfield  {title} {\bibinfo {title} {Identifying quantum phase transitions using artificial neural networks on experimental data},\ }\href {https://doi.org/10.1038/s41567-019-0554-0} {\bibfield  {journal} {\bibinfo  {journal} {Nat. Phys.}\ }\textbf {\bibinfo {volume} {15}},\ \bibinfo {pages} {917} (\bibinfo {year} {2019})}\BibitemShut {NoStop}%
\bibitem [{\citenamefont {Dong}\ \emph {et~al.}(2019)\citenamefont {Dong}, \citenamefont {Pollmann},\ and\ \citenamefont {Zhang}}]{paper_SL_8}%
  \BibitemOpen
  \bibfield  {author} {\bibinfo {author} {\bibfnamefont {X.-Y.}\ \bibnamefont {Dong}}, \bibinfo {author} {\bibfnamefont {F.}~\bibnamefont {Pollmann}},\ and\ \bibinfo {author} {\bibfnamefont {X.-F.}\ \bibnamefont {Zhang}},\ }\bibfield  {title} {\bibinfo {title} {Machine learning of quantum phase transitions},\ }\href {https://doi.org/10.1103/PhysRevB.99.121104} {\bibfield  {journal} {\bibinfo  {journal} {Phys. Rev. B}\ }\textbf {\bibinfo {volume} {99}},\ \bibinfo {pages} {121104(R)} (\bibinfo {year} {2019})}\BibitemShut {NoStop}%
\bibitem [{\citenamefont {Wang}(2016)}]{paper_USL_1}%
  \BibitemOpen
  \bibfield  {author} {\bibinfo {author} {\bibfnamefont {L.}~\bibnamefont {Wang}},\ }\bibfield  {title} {\bibinfo {title} {Discovering phase transitions with unsupervised learning},\ }\href {https://doi.org/10.1103/PhysRevB.94.195105} {\bibfield  {journal} {\bibinfo  {journal} {Phys. Rev. B}\ }\textbf {\bibinfo {volume} {94}},\ \bibinfo {pages} {195105} (\bibinfo {year} {2016})}\BibitemShut {NoStop}%
\bibitem [{\citenamefont {Wetzel}(2017)}]{paper_USL_2}%
  \BibitemOpen
  \bibfield  {author} {\bibinfo {author} {\bibfnamefont {S.~J.}\ \bibnamefont {Wetzel}},\ }\bibfield  {title} {\bibinfo {title} {Unsupervised learning of phase transitions: From principal component analysis to variational autoencoders},\ }\href {https://doi.org/10.1103/PhysRevE.96.022140} {\bibfield  {journal} {\bibinfo  {journal} {Phys. Rev. E}\ }\textbf {\bibinfo {volume} {96}},\ \bibinfo {pages} {022140} (\bibinfo {year} {2017})}\BibitemShut {NoStop}%
\bibitem [{\citenamefont {Ch'ng}\ \emph {et~al.}(2018)\citenamefont {Ch'ng}, \citenamefont {Vazquez},\ and\ \citenamefont {Khatami}}]{paper_USL_3}%
  \BibitemOpen
  \bibfield  {author} {\bibinfo {author} {\bibfnamefont {K.}~\bibnamefont {Ch'ng}}, \bibinfo {author} {\bibfnamefont {N.}~\bibnamefont {Vazquez}},\ and\ \bibinfo {author} {\bibfnamefont {E.}~\bibnamefont {Khatami}},\ }\bibfield  {title} {\bibinfo {title} {Unsupervised machine learning account of magnetic transitions in the hubbard model},\ }\href {https://doi.org/10.1103/PhysRevE.97.013306} {\bibfield  {journal} {\bibinfo  {journal} {Phys. Rev. E}\ }\textbf {\bibinfo {volume} {97}},\ \bibinfo {pages} {013306} (\bibinfo {year} {2018})}\BibitemShut {NoStop}%
\bibitem [{\citenamefont {Liu}\ and\ \citenamefont {van Nieuwenburg}(2018)}]{paper_USL_4}%
  \BibitemOpen
  \bibfield  {author} {\bibinfo {author} {\bibfnamefont {Y.-H.}\ \bibnamefont {Liu}}\ and\ \bibinfo {author} {\bibfnamefont {E.~P.~L.}\ \bibnamefont {van Nieuwenburg}},\ }\bibfield  {title} {\bibinfo {title} {Discriminative cooperative networks for detecting phase transitions},\ }\href {https://doi.org/10.1103/PhysRevLett.120.176401} {\bibfield  {journal} {\bibinfo  {journal} {Phys. Rev. Lett.}\ }\textbf {\bibinfo {volume} {120}},\ \bibinfo {pages} {176401} (\bibinfo {year} {2018})}\BibitemShut {NoStop}%
\bibitem [{\citenamefont {Sch\"afer}\ and\ \citenamefont {L\"orch}(2019)}]{paper_USL_5}%
  \BibitemOpen
  \bibfield  {author} {\bibinfo {author} {\bibfnamefont {F.}~\bibnamefont {Sch\"afer}}\ and\ \bibinfo {author} {\bibfnamefont {N.}~\bibnamefont {L\"orch}},\ }\bibfield  {title} {\bibinfo {title} {Vector field divergence of predictive model output as indication of phase transitions},\ }\href {https://doi.org/10.1103/PhysRevE.99.062107} {\bibfield  {journal} {\bibinfo  {journal} {Phys. Rev. E}\ }\textbf {\bibinfo {volume} {99}},\ \bibinfo {pages} {062107} (\bibinfo {year} {2019})}\BibitemShut {NoStop}%
\bibitem [{\citenamefont {Canabarro}\ \emph {et~al.}(2019)\citenamefont {Canabarro}, \citenamefont {Fanchini}, \citenamefont {Malvezzi}, \citenamefont {Pereira},\ and\ \citenamefont {Chaves}}]{paper_USL_6}%
  \BibitemOpen
  \bibfield  {author} {\bibinfo {author} {\bibfnamefont {A.}~\bibnamefont {Canabarro}}, \bibinfo {author} {\bibfnamefont {F.~F.}\ \bibnamefont {Fanchini}}, \bibinfo {author} {\bibfnamefont {A.~L.}\ \bibnamefont {Malvezzi}}, \bibinfo {author} {\bibfnamefont {R.}~\bibnamefont {Pereira}},\ and\ \bibinfo {author} {\bibfnamefont {R.}~\bibnamefont {Chaves}},\ }\bibfield  {title} {\bibinfo {title} {Unveiling phase transitions with machine learning},\ }\href {https://doi.org/10.1103/PhysRevB.100.045129} {\bibfield  {journal} {\bibinfo  {journal} {Phys. Rev. B}\ }\textbf {\bibinfo {volume} {100}},\ \bibinfo {pages} {045129} (\bibinfo {year} {2019})}\BibitemShut {NoStop}%
\bibitem [{\citenamefont {Rodriguez-Nieva}\ and\ \citenamefont {Scheurer}(2019)}]{paper_USL_7}%
  \BibitemOpen
  \bibfield  {author} {\bibinfo {author} {\bibfnamefont {J.~F.}\ \bibnamefont {Rodriguez-Nieva}}\ and\ \bibinfo {author} {\bibfnamefont {M.~S.}\ \bibnamefont {Scheurer}},\ }\bibfield  {title} {\bibinfo {title} {Identifying topological order through unsupervised machine learning},\ }\href {https://doi.org/10.1038/s41567-019-0512-x} {\bibfield  {journal} {\bibinfo  {journal} {Nat. Phys.}\ }\textbf {\bibinfo {volume} {15}},\ \bibinfo {pages} {790–795} (\bibinfo {year} {2019})}\BibitemShut {NoStop}%
\bibitem [{\citenamefont {Walker}\ \emph {et~al.}(2020)\citenamefont {Walker}, \citenamefont {Tam},\ and\ \citenamefont {Jarrell}}]{paper_USL_8}%
  \BibitemOpen
  \bibfield  {author} {\bibinfo {author} {\bibfnamefont {N.}~\bibnamefont {Walker}}, \bibinfo {author} {\bibfnamefont {K.-M.}\ \bibnamefont {Tam}},\ and\ \bibinfo {author} {\bibfnamefont {M.}~\bibnamefont {Jarrell}},\ }\bibfield  {title} {\bibinfo {title} {Deep learning on the 2-dimensional ising model to extract the crossover region with a variational autoencoder},\ }\href {https://doi.org/10.1038/s41598-020-69848-5} {\bibfield  {journal} {\bibinfo  {journal} {Sci Rep}\ }\textbf {\bibinfo {volume} {10}},\ \bibinfo {pages} {13047} (\bibinfo {year} {2020})}\BibitemShut {NoStop}%
\bibitem [{\citenamefont {Scheurer}\ and\ \citenamefont {Slager}(2020)}]{paper_USL_9}%
  \BibitemOpen
  \bibfield  {author} {\bibinfo {author} {\bibfnamefont {M.~S.}\ \bibnamefont {Scheurer}}\ and\ \bibinfo {author} {\bibfnamefont {R.-J.}\ \bibnamefont {Slager}},\ }\bibfield  {title} {\bibinfo {title} {Unsupervised machine learning and band topology},\ }\href {https://doi.org/10.1103/PhysRevLett.124.226401} {\bibfield  {journal} {\bibinfo  {journal} {Phys. Rev. Lett.}\ }\textbf {\bibinfo {volume} {124}},\ \bibinfo {pages} {226401} (\bibinfo {year} {2020})}\BibitemShut {NoStop}%
\bibitem [{\citenamefont {Wang}\ \emph {et~al.}(2021)\citenamefont {Wang}, \citenamefont {Zhang}, \citenamefont {Hua},\ and\ \citenamefont {Wei}}]{paper_USL_10}%
  \BibitemOpen
  \bibfield  {author} {\bibinfo {author} {\bibfnamefont {J.}~\bibnamefont {Wang}}, \bibinfo {author} {\bibfnamefont {W.}~\bibnamefont {Zhang}}, \bibinfo {author} {\bibfnamefont {T.}~\bibnamefont {Hua}},\ and\ \bibinfo {author} {\bibfnamefont {T.-C.}\ \bibnamefont {Wei}},\ }\bibfield  {title} {\bibinfo {title} {Unsupervised learning of topological phase transitions using the calinski-harabaz index},\ }\href {https://doi.org/10.1103/PhysRevResearch.3.013074} {\bibfield  {journal} {\bibinfo  {journal} {Phys. Rev. Res.}\ }\textbf {\bibinfo {volume} {3}},\ \bibinfo {pages} {013074} (\bibinfo {year} {2021})}\BibitemShut {NoStop}%
\bibitem [{\citenamefont {Mendes-Santos}\ \emph {et~al.}(2021{\natexlab{a}})\citenamefont {Mendes-Santos}, \citenamefont {Angelone}, \citenamefont {Rodriguez}, \citenamefont {Fazio},\ and\ \citenamefont {Dalmonte}}]{paper_USL_11}%
  \BibitemOpen
  \bibfield  {author} {\bibinfo {author} {\bibfnamefont {T.}~\bibnamefont {Mendes-Santos}}, \bibinfo {author} {\bibfnamefont {A.}~\bibnamefont {Angelone}}, \bibinfo {author} {\bibfnamefont {A.}~\bibnamefont {Rodriguez}}, \bibinfo {author} {\bibfnamefont {R.}~\bibnamefont {Fazio}},\ and\ \bibinfo {author} {\bibfnamefont {M.}~\bibnamefont {Dalmonte}},\ }\bibfield  {title} {\bibinfo {title} {Intrinsic dimension of path integrals: Data-mining quantum criticality and emergent simplicity},\ }\href {https://doi.org/10.1103/PRXQuantum.2.030332} {\bibfield  {journal} {\bibinfo  {journal} {PRX Quantum}\ }\textbf {\bibinfo {volume} {2}},\ \bibinfo {pages} {030332} (\bibinfo {year} {2021}{\natexlab{a}})}\BibitemShut {NoStop}%
\bibitem [{\citenamefont {K{\"a}ming}\ \emph {et~al.}(2021)\citenamefont {K{\"a}ming}, \citenamefont {Dawid}, \citenamefont {Kottmann}, \citenamefont {Lewenstein}, \citenamefont {Sengstock}, \citenamefont {Dauphin},\ and\ \citenamefont {Weitenberg}}]{paper_USL_12}%
  \BibitemOpen
  \bibfield  {author} {\bibinfo {author} {\bibfnamefont {N.}~\bibnamefont {K{\"a}ming}}, \bibinfo {author} {\bibfnamefont {A.}~\bibnamefont {Dawid}}, \bibinfo {author} {\bibfnamefont {K.}~\bibnamefont {Kottmann}}, \bibinfo {author} {\bibfnamefont {M.}~\bibnamefont {Lewenstein}}, \bibinfo {author} {\bibfnamefont {K.}~\bibnamefont {Sengstock}}, \bibinfo {author} {\bibfnamefont {A.}~\bibnamefont {Dauphin}},\ and\ \bibinfo {author} {\bibfnamefont {C.}~\bibnamefont {Weitenberg}},\ }\bibfield  {title} {\bibinfo {title} {Unsupervised machine learning of topological phase transitions from experimental data},\ }\href {https://doi.org/10.1088/2632-2153/abffe7} {\bibfield  {journal} {\bibinfo  {journal} {Mach. Learn.: Sci. Technol.}\ }\textbf {\bibinfo {volume} {2}},\ \bibinfo {pages} {035037} (\bibinfo {year} {2021})}\BibitemShut {NoStop}%
\bibitem [{\citenamefont {Tirelli}\ \emph {et~al.}(2022)\citenamefont {Tirelli}, \citenamefont {Carvalho}, \citenamefont {Oliveira}, \citenamefont {de~Lima}, \citenamefont {Costa},\ and\ \citenamefont {dos Santos}}]{paper_USL_13}%
  \BibitemOpen
  \bibfield  {author} {\bibinfo {author} {\bibfnamefont {A.}~\bibnamefont {Tirelli}}, \bibinfo {author} {\bibfnamefont {D.~O.}\ \bibnamefont {Carvalho}}, \bibinfo {author} {\bibfnamefont {L.~A.}\ \bibnamefont {Oliveira}}, \bibinfo {author} {\bibfnamefont {J.~P.}\ \bibnamefont {de~Lima}}, \bibinfo {author} {\bibfnamefont {N.~C.}\ \bibnamefont {Costa}},\ and\ \bibinfo {author} {\bibfnamefont {R.~R.}\ \bibnamefont {dos Santos}},\ }\bibfield  {title} {\bibinfo {title} {Unsupervised machine learning approaches to the $q$-state potts model},\ }\href {https://doi.org/10.1140/epjb/s10051-022-00453-3} {\bibfield  {journal} {\bibinfo  {journal} {Eur. Phys. J. B}\ }\textbf {\bibinfo {volume} {95}},\ \bibinfo {pages} {013306} (\bibinfo {year} {2022})}\BibitemShut {NoStop}%
\bibitem [{\citenamefont {Ng}\ and\ \citenamefont {Yang}(2023)}]{paper_USL_14}%
  \BibitemOpen
  \bibfield  {author} {\bibinfo {author} {\bibfnamefont {K.-K.}\ \bibnamefont {Ng}}\ and\ \bibinfo {author} {\bibfnamefont {M.-F.}\ \bibnamefont {Yang}},\ }\bibfield  {title} {\bibinfo {title} {Unsupervised learning of phase transitions via modified anomaly detection with autoencoders},\ }\href {https://doi.org/10.1103/PhysRevB.108.214428} {\bibfield  {journal} {\bibinfo  {journal} {Phys. Rev. B}\ }\textbf {\bibinfo {volume} {108}},\ \bibinfo {pages} {214428} (\bibinfo {year} {2023})}\BibitemShut {NoStop}%
\bibitem [{\citenamefont {Sadoune}\ \emph {et~al.}(2023)\citenamefont {Sadoune}, \citenamefont {Giudici}, \citenamefont {Liu},\ and\ \citenamefont {Pollet}}]{paper_USL_15}%
  \BibitemOpen
  \bibfield  {author} {\bibinfo {author} {\bibfnamefont {N.}~\bibnamefont {Sadoune}}, \bibinfo {author} {\bibfnamefont {G.}~\bibnamefont {Giudici}}, \bibinfo {author} {\bibfnamefont {K.}~\bibnamefont {Liu}},\ and\ \bibinfo {author} {\bibfnamefont {L.}~\bibnamefont {Pollet}},\ }\bibfield  {title} {\bibinfo {title} {Unsupervised interpretable learning of phases from many-qubit systems},\ }\href {https://doi.org/10.1103/PhysRevResearch.5.013082} {\bibfield  {journal} {\bibinfo  {journal} {Phys. Rev. Res.}\ }\textbf {\bibinfo {volume} {5}},\ \bibinfo {pages} {013082} (\bibinfo {year} {2023})}\BibitemShut {NoStop}%
\bibitem [{\citenamefont {Mendes-Santos}\ \emph {et~al.}(2021{\natexlab{b}})\citenamefont {Mendes-Santos}, \citenamefont {Turkeshi}, \citenamefont {Dalmonte},\ and\ \citenamefont {Rodriguez}}]{paper_USL_16}%
  \BibitemOpen
  \bibfield  {author} {\bibinfo {author} {\bibfnamefont {T.}~\bibnamefont {Mendes-Santos}}, \bibinfo {author} {\bibfnamefont {X.}~\bibnamefont {Turkeshi}}, \bibinfo {author} {\bibfnamefont {M.}~\bibnamefont {Dalmonte}},\ and\ \bibinfo {author} {\bibfnamefont {A.}~\bibnamefont {Rodriguez}},\ }\bibfield  {title} {\bibinfo {title} {Unsupervised learning universal critical behavior via the intrinsic dimension},\ }\href {https://doi.org/10.1103/PhysRevX.11.011040} {\bibfield  {journal} {\bibinfo  {journal} {Phys. Rev. X}\ }\textbf {\bibinfo {volume} {11}},\ \bibinfo {pages} {011040} (\bibinfo {year} {2021}{\natexlab{b}})}\BibitemShut {NoStop}%
\bibitem [{\citenamefont {Ho}\ and\ \citenamefont {Wang}(2021)}]{paper_Ho_1}%
  \BibitemOpen
  \bibfield  {author} {\bibinfo {author} {\bibfnamefont {C.-T.}\ \bibnamefont {Ho}}\ and\ \bibinfo {author} {\bibfnamefont {D.-W.}\ \bibnamefont {Wang}},\ }\bibfield  {title} {\bibinfo {title} {Robust identification of topological phase transition by self-supervised machine learning approach},\ }\href {https://doi.org/10.1088/1367-2630/ac1709} {\bibfield  {journal} {\bibinfo  {journal} {New J. Phys.}\ }\textbf {\bibinfo {volume} {23}},\ \bibinfo {pages} {083021} (\bibinfo {year} {2021})}\BibitemShut {NoStop}%
\bibitem [{\citenamefont {Ho}\ and\ \citenamefont {Wang}(2023)}]{paper_Ho_2}%
  \BibitemOpen
  \bibfield  {author} {\bibinfo {author} {\bibfnamefont {C.-T.}\ \bibnamefont {Ho}}\ and\ \bibinfo {author} {\bibfnamefont {D.-W.}\ \bibnamefont {Wang}},\ }\bibfield  {title} {\bibinfo {title} {Self-supervised ensemble learning: A universal method for phase transition classification of many-body systems},\ }\href {https://doi.org/10.1103/PhysRevResearch.5.043090} {\bibfield  {journal} {\bibinfo  {journal} {Phys. Rev. Res.}\ }\textbf {\bibinfo {volume} {5}},\ \bibinfo {pages} {043090} (\bibinfo {year} {2023})}\BibitemShut {NoStop}%
\bibitem [{\citenamefont {F.R.S.}(1901)}]{paper_PCA}%
  \BibitemOpen
  \bibfield  {author} {\bibinfo {author} {\bibfnamefont {K.~P.}\ \bibnamefont {F.R.S.}},\ }\bibfield  {title} {\bibinfo {title} {Liii. on lines and planes of closest fit to systems of points in space},\ }\href {https://doi.org/10.1080/14786440109462720} {\bibfield  {journal} {\bibinfo  {journal} {The London, Edinburgh, and Dublin Philosophical Magazine and Journal of Science}\ }\textbf {\bibinfo {volume} {2}},\ \bibinfo {pages} {559} (\bibinfo {year} {1901})}\BibitemShut {NoStop}%
\bibitem [{\citenamefont {Tantivasadakarn}\ \emph {et~al.}(2023)\citenamefont {Tantivasadakarn}, \citenamefont {Thorngren}, \citenamefont {Vishwanath},\ and\ \citenamefont {Verresen}}]{paper_cluster-ising_1}%
  \BibitemOpen
  \bibfield  {author} {\bibinfo {author} {\bibfnamefont {N.}~\bibnamefont {Tantivasadakarn}}, \bibinfo {author} {\bibfnamefont {R.}~\bibnamefont {Thorngren}}, \bibinfo {author} {\bibfnamefont {A.}~\bibnamefont {Vishwanath}},\ and\ \bibinfo {author} {\bibfnamefont {R.}~\bibnamefont {Verresen}},\ }\bibfield  {title} {\bibinfo {title} {Pivot hamiltonians as generators of symmetry and entanglement},\ }\href {https://doi.org/10.21468/SciPostPhys.14.2.012} {\bibfield  {journal} {\bibinfo  {journal} {Sci. Rep.}\ }\textbf {\bibinfo {volume} {14}},\ \bibinfo {pages} {012} (\bibinfo {year} {2023})}\BibitemShut {NoStop}%
\bibitem [{\citenamefont {Choi}\ \emph {et~al.}(2024)\citenamefont {Choi}, \citenamefont {Knap},\ and\ \citenamefont {Pollmann}}]{paper_cluster-ising_2}%
  \BibitemOpen
  \bibfield  {author} {\bibinfo {author} {\bibfnamefont {W.}~\bibnamefont {Choi}}, \bibinfo {author} {\bibfnamefont {M.}~\bibnamefont {Knap}},\ and\ \bibinfo {author} {\bibfnamefont {F.}~\bibnamefont {Pollmann}},\ }\bibfield  {title} {\bibinfo {title} {Finite-temperature entanglement negativity of fermionic symmetry-protected topological phases and quantum critical points in one dimension},\ }\href {https://doi.org/10.1103/PhysRevB.109.115132} {\bibfield  {journal} {\bibinfo  {journal} {Phys. Rev. B}\ }\textbf {\bibinfo {volume} {109}},\ \bibinfo {pages} {115132} (\bibinfo {year} {2024})}\BibitemShut {NoStop}%
\bibitem [{\citenamefont {Yu-Chin}\ \emph {et~al.}(2023)\citenamefont {Yu-Chin}, \citenamefont {Li}, \citenamefont {Ming-Chiang}, \citenamefont {Amico},\ and\ \citenamefont {Kwek}}]{paper_alternating-xxz_1}%
  \BibitemOpen
  \bibfield  {author} {\bibinfo {author} {\bibfnamefont {T.}~\bibnamefont {Yu-Chin}}, \bibinfo {author} {\bibfnamefont {D.}~\bibnamefont {Li}}, \bibinfo {author} {\bibfnamefont {C.}~\bibnamefont {Ming-Chiang}}, \bibinfo {author} {\bibfnamefont {L.}~\bibnamefont {Amico}},\ and\ \bibinfo {author} {\bibfnamefont {L.-C.}\ \bibnamefont {Kwek}},\ }\bibfield  {title} {\bibinfo {title} {Entanglement convertibility by sweeping through the quantum phases of the alternating bonds xxz chain},\ }\href {https://doi.org/10.21468/SciPostPhys.14.2.012} {\bibfield  {journal} {\bibinfo  {journal} {Sci. Rep.}\ }\textbf {\bibinfo {volume} {14}},\ \bibinfo {pages} {012} (\bibinfo {year} {2023})}\BibitemShut {NoStop}%
\bibitem [{\citenamefont {Kitaev}(2006)}]{paper_2d-kitaev}%
  \BibitemOpen
  \bibfield  {author} {\bibinfo {author} {\bibfnamefont {A.}~\bibnamefont {Kitaev}},\ }\bibfield  {title} {\bibinfo {title} {Anyons in an exactly solved model and beyond},\ }\href {https://doi.org/https://doi.org/10.1016/j.aop.2005.10.005} {\bibfield  {journal} {\bibinfo  {journal} {Ann. Phys}\ }\textbf {\bibinfo {volume} {321}},\ \bibinfo {pages} {2} (\bibinfo {year} {2006})}\BibitemShut {NoStop}%
\bibitem [{\citenamefont {Sachdev}(1999)}]{book_sachdev}%
  \BibitemOpen
  \bibfield  {author} {\bibinfo {author} {\bibfnamefont {S.}~\bibnamefont {Sachdev}},\ }\bibfield  {title} {\bibinfo {title} {Quantum phase transitions},\ }\href {https://doi.org/10.1088/2058-7058/12/4/23} {\bibfield  {journal} {\bibinfo  {journal} {Physics world}\ }\textbf {\bibinfo {volume} {12}},\ \bibinfo {pages} {33} (\bibinfo {year} {1999})}\BibitemShut {NoStop}%
\bibitem [{\citenamefont {Giamarchi}(2003)}]{book_giamarchi}%
  \BibitemOpen
  \bibfield  {author} {\bibinfo {author} {\bibfnamefont {T.}~\bibnamefont {Giamarchi}},\ }\href@noop {} {\emph {\bibinfo {title} {Quantum physics in one dimension}}},\ Vol.\ \bibinfo {volume} {121}\ (\bibinfo  {publisher} {Clarendon press},\ \bibinfo {year} {2003})\BibitemShut {NoStop}%
\bibitem [{\citenamefont {Su}\ \emph {et~al.}(1979)\citenamefont {Su}, \citenamefont {Schrieffer},\ and\ \citenamefont {Heeger}}]{paper_SSH}%
  \BibitemOpen
  \bibfield  {author} {\bibinfo {author} {\bibfnamefont {W.~P.}\ \bibnamefont {Su}}, \bibinfo {author} {\bibfnamefont {J.~R.}\ \bibnamefont {Schrieffer}},\ and\ \bibinfo {author} {\bibfnamefont {A.~J.}\ \bibnamefont {Heeger}},\ }\bibfield  {title} {\bibinfo {title} {Solitons in polyacetylene},\ }\href {https://doi.org/10.1103/PhysRevLett.42.1698} {\bibfield  {journal} {\bibinfo  {journal} {Phys. Rev. Lett.}\ }\textbf {\bibinfo {volume} {42}},\ \bibinfo {pages} {1698} (\bibinfo {year} {1979})}\BibitemShut {NoStop}%
\bibitem [{\citenamefont {Friedman}(1978)}]{paper_2d-tfim}%
  \BibitemOpen
  \bibfield  {author} {\bibinfo {author} {\bibfnamefont {Z.}~\bibnamefont {Friedman}},\ }\bibfield  {title} {\bibinfo {title} {Ising model with a transverse field in two dimensions: Phase diagram and critical properties from a real-space renormalization group},\ }\href {https://doi.org/10.1103/PhysRevB.17.1429} {\bibfield  {journal} {\bibinfo  {journal} {Phys. Rev. B}\ }\textbf {\bibinfo {volume} {17}},\ \bibinfo {pages} {1429} (\bibinfo {year} {1978})}\BibitemShut {NoStop}%
\bibitem [{\citenamefont {Fishman}\ \emph {et~al.}(2022{\natexlab{a}})\citenamefont {Fishman}, \citenamefont {White},\ and\ \citenamefont {Stoudenmire}}]{ITensor}%
  \BibitemOpen
  \bibfield  {author} {\bibinfo {author} {\bibfnamefont {M.}~\bibnamefont {Fishman}}, \bibinfo {author} {\bibfnamefont {S.~R.}\ \bibnamefont {White}},\ and\ \bibinfo {author} {\bibfnamefont {E.~M.}\ \bibnamefont {Stoudenmire}},\ }\bibfield  {title} {\bibinfo {title} {The itensor software library for tensor network calculations},\ }\href {https://doi.org/10.21468/SciPostPhysCodeb.4} {\bibfield  {journal} {\bibinfo  {journal} {SciPost Phys. Codebases}\ ,\ \bibinfo {pages} {4}} (\bibinfo {year} {2022}{\natexlab{a}})}\BibitemShut {NoStop}%
\bibitem [{\citenamefont {Fishman}\ \emph {et~al.}(2022{\natexlab{b}})\citenamefont {Fishman}, \citenamefont {White},\ and\ \citenamefont {Stoudenmire}}]{ITensor-r0.3}%
  \BibitemOpen
  \bibfield  {author} {\bibinfo {author} {\bibfnamefont {M.}~\bibnamefont {Fishman}}, \bibinfo {author} {\bibfnamefont {S.~R.}\ \bibnamefont {White}},\ and\ \bibinfo {author} {\bibfnamefont {E.~M.}\ \bibnamefont {Stoudenmire}},\ }\bibfield  {title} {\bibinfo {title} {Codebase release 0.3 for itensor},\ }\href {https://doi.org/10.21468/SciPostPhysCodeb.4-r0.3} {\bibfield  {journal} {\bibinfo  {journal} {SciPost Phys. Codebases}\ ,\ \bibinfo {pages} {4}} (\bibinfo {year} {2022}{\natexlab{b}})}\BibitemShut {NoStop}%
\end{thebibliography}%


\end{document}